\documentclass[11pt,twoside,english]{article}
\usepackage[]{fontenc}
\usepackage[latin9]{inputenc}
\usepackage{geometry}
\usepackage{bm}
\usepackage{amsmath}
\usepackage{setspace}

\makeatletter

\providecommand{\tabularnewline}{\\}

 \makeatletter
\@addtoreset{equation}{section}

\makeatother

\makeatother

\usepackage{babel}

\begin{document}
\setlength{\baselineskip}{16pt}

\title{{\huge Liouville Equation in }1/8{\huge{} BPS Geometries\vspace{0.3in}
}}

\author{{\Large Yoshihiro Mitsuka}%
\thanks{mitsuka@hep1.c.u-tokyo.ac.jp%
}}

\date{}

\maketitle
\begin{center}
\emph{\large Institute of Physics, University of Tokyo,}\\
\emph{\large Komaba, Meguro-ku, Tokyo 153-8902 Japan}\emph{\vspace{1in}
}
\par\end{center}

\begin{abstract}
\begin{onehalfspace}
We {\normalsize investigate the $\frac{1}{8}$ BPS geometries with
$SU(2)\times U(1)\times SO(4)\times R$ symmetry in IIB supergravity
which were classified by Gava et al, (hep-th/0611065). It is desirable
to have a complete set of differential equations imposed on the controlling
functions such that they are not only necessary but also sufficient
to produce supergravity solutions with those symmetries. We work on
this issue and find a new differential equation for the controlling
functions. For a special case, we exhaust all the remaining constraints
and show that they reduce to one Liouville equation. The solutions
of this equation produce geometries which are locally equivalent to
the near horizon geometries of intersecting D3-branes. }{\normalsize \par}
\end{onehalfspace}

\emph{\vspace{0.8in}
}
\end{abstract}
UT-Komaba/08-9

\thispagestyle{empty}\setcounter{page}{0} 

\clearpage{}

\section{Introduction}

In studying $AdS/CFT$ correspondence, it is an interesting subject
to examine the duality at the regions in which the state is highly
excited to the extent that the backreactions in the gravity side are
not negligible. In this sense, recent developments in the analysis
of BPS geometries in supergravity theories are important as possible
sources of information, and those results deserve to be studied in
more details. In \cite{Lin:2004nb}, by analyzing the BPS condition
in IIB supergravity, a class of $\frac{1}{2}$ BPS geometries with
$SO(4)\times SO(4)\times R$ symmetry were concisely written in terms
of one function on a three-dimensional subspace and one differential
equation imposed on that function was obtained so that the geometries
were classified by the boundary conditions on a two-dimensional plane. 

After this work, several works have been done to treat more general
situations (\cite{Kim:2005ez,Donos:2006iy,Donos:2006ms,Gava:2006pu,Chen:2007du,Lunin:2008tf}).
Among them we concentrate on the result of \cite{Gava:2006pu}, in
which the case of $SU(2)\times U(1)\times SO(4)\times R$ symmetry
was studied and as a result, a class of $\frac{1}{8}$ BPS geometries
were written in terms of four functions and four differential equations
for them have been found. One of the tasks left to be done is to exhaust
the constraints for the controlling functions coming from the BPS
condition and the equations of motion so that they form a framework
to produce solutions of the supergravity with the above symmetries.
Another is to find implications for the dual field theories which
may arise as a result of these analyses on the gravity sides. 

In this paper, to contribute in these directions, we report some new
facts about the geometries considered in \cite{Gava:2006pu}. First
we find that the differential equations obtained in \cite{Gava:2006pu}
are not sufficient to exclude all the geometries which does not solve
the supergravity equations of motion and present an additional differential
equation which should be imposed on the four controlling functions.
Second we find a restricted class of geometries in which the four
functions and the five differential equations reduce to two functions
and two differential equations. We pick up all the remaining constraints
imposed by the BPS condition and the equations of motion for this
class of geometries and find that one of the two controlling functions
must be constant. The differential equation imposed on the remaining
function becomes a Liouville equation having its cosmological constant
as a free parameter and all the geometries which correspond to solutions
of that equation are locally equivalent to the near horizon geometries
of intersecting D3-brane systems. Thus one of the above mentioned
tasks is completed in this restricted case. In this second part, the
roles of the new differential equation are very crucial. We also argue
on the T-duality transformation to D1-D5 system and possible future
directions. 

Apart from the discovery of the new differential equation for the
general geometries in the first part, the restriction we consider
in the second part eliminates perhaps most informative geometries,
that is, asymptotically $AdS_{5}\times S^{5}$ geometries. Nevertheless
we consider that the appearance of geometries with another asymptotics
should be considered as an important property because in some sense
it relates two class of geometries with different asymptotics. If
this relation is interpreted as a relation between the dual CFTs,
we obtain a strong support for $AdS/CFT$ correspondence in backreacted
region. 

This paper is organized as follows. In Section \ref{sec:1/8-BPS-geometries}
we review the analysis of \cite{Gava:2006pu} and explain how the
new differential equation appears. In Section \ref{sec:s3xs3} we
take a limit which reduces the expressions for the geometries to simple
forms, exhaust the constraints for them and exhibit the roles of the
new differential equation. In Section \ref{sec:Conclusion} we discuss
the possibilities for applying our result.

\section{1/8 BPS geometries with $SU(2)\times U(1)\times SO(4)\times R$\label{sec:1/8-BPS-geometries}}

The purpose of this section is to examine the result of \cite{Gava:2006pu}
and point out the existence of an additional constraint \eqref{eq:consistency5}.
We start with a review of the analysis in \cite{Gava:2006pu}, as
the derivation of \eqref{eq:consistency5} is related to its details.

\subsubsection*{\emph{Setup}}

In \cite{Lin:2004nb}, a class of type IIB $\frac{1}{2}$ BPS geometries
consisting of the metric and five-from flux with $SO(4)\times SO(4)\times R$
symmetry has been obtained through the procedure in which two $S^{3}$s
were set in the starting ansatz and the Killing spinor equation was
analyzed leading to the result that a timelike Killing vector was
found and constraints for the other components of the geometry were
picked up. In \cite{Gava:2006pu}, that analysis was extended to $SU(2)\times U(1)\times SO(4)\times R$
case. The basic idea is that we replace one of the $S^{3}$s in \cite{Lin:2004nb}
with a squashed $S^{3}$ to break the $SU(2)_{R}$ in the isometry
group $SO(4)=SU(2)_{L}\times SU(2)_{R}$ of $S^{3}$. The ansatz for
the $SU(2)_{L}\times U(1)\times SO(4)$ symmetric metric and five-form
flux is given by\begin{align}
ds^{2} & =g_{\mu\nu}dx^{\mu}dx^{\nu}+\rho_{1}^{2}\left[\sigma_{\hat{1}}^{2}+\sigma_{\hat{2}}^{2}\right]+\rho_{3}^{2}\left(\sigma_{\hat{3}}-A_{\mu}dx^{\mu}\right)^{2}+\tilde{\rho}^{2}d\tilde{\Omega}_{3}^{2}\label{eq:ansatz-met}\\
F_{5} & =-\left(G_{\bar{\mu}\bar{\nu}}e^{\bar{\mu}}\wedge e^{\bar{\nu}}\wedge e^{\bar{\hat{1}}}\wedge e^{\bar{\hat{2}}}\wedge e^{\bar{\hat{3}}}+*_{4}\tilde{V}\wedge e^{\bar{\hat{1}}}\wedge e^{\bar{\hat{2}}}+*_{4}\tilde{g}\wedge e^{\bar{\hat{3}}}\right)\nonumber \\
 & +\left(\tilde{G}_{\bar{\mu}\bar{\nu}}e^{\bar{\mu}}\wedge e^{\bar{\nu}}+\tilde{V}_{\bar{\mu}}e^{\bar{\mu}}\wedge e^{\bar{\hat{3}}}+\tilde{g}e^{\bar{\hat{1}}}\wedge e^{\bar{\hat{2}}}\right)\wedge\tilde{\rho}^{3}d\tilde{\Omega}_{3}.\label{eq:ansatz-flux}\end{align}
Here $\mu,\nu$ take values 0,1,2,3, and $g_{\mu\nu},\rho_{1},\rho_{3},\tilde{\rho},A_{\mu},G_{\mu\nu},\tilde{G}_{\mu\nu},\tilde{V},\tilde{g}$
depend only on the four-dimensional coordinate $x^{\mu}$. $\sigma_{\hat{i}}$s
are the left-invariant 1-forms used for building the metrics of squashed
three-spheres. The explicit forms of them are\begin{align}
\sigma_{\hat{1}} & =-\frac{1}{2}\left(\cos\hat{\psi}d\hat{\theta}+\sin\hat{\psi}\sin\hat{\theta}d\hat{\phi}\right)\nonumber \\
\sigma_{\hat{2}} & =-\frac{1}{2}\left(-\sin\hat{\psi}d\hat{\theta}+\cos\hat{\psi}\sin\hat{\theta}d\hat{\phi}\right)\nonumber \\
\sigma_{\hat{3}} & =-\frac{1}{2}\left(d\hat{\psi}+\cos\hat{\theta}d\hat{\phi}\right)\label{eq:explicit-forms-of-squashed-s3}\end{align}
 (see appendix \ref{sec:Squashed-three-sphere} for notations related
to the symmetry). $e^{\bar{\mu},\bar{\nu}},e^{\bar{\hat{1}},\bar{\hat{2}},\bar{\hat{3}}}$
are the vierbein 1-forms with their indices in the respective tangent
subspaces. We take $e^{\bar{\hat{1}},\bar{\hat{2}},\bar{\hat{3}}}$
to be of the forms associated with $\sigma_{\hat{1},\hat{2},\hat{3}}$
:\[
e^{\bar{\hat{1}},\bar{\hat{2}}}=\rho_{1}\sigma_{\hat{1},\hat{2}},\quad e^{\bar{\hat{3}}}=\rho_{3}\left(\sigma_{\hat{3}}-A\right).\]
$*_{4}$ is the Hodge dual in the four-dimensional subspaces described
by the first term in the metric. $d\tilde{\Omega}_{3}^{2}$ is a metric
of $S^{3}$ and $d\tilde{\Omega}_{3}$ is its volume form. Because
we have set the coefficient of $\sigma_{\hat{1}}^{2},\sigma_{\hat{2}}^{2}$
equal, the translation of $\hat{\psi}$ gives the extra $U(1)$ symmetry.
The five-form $F_{5}$ must satisfy two constraints. One is the self-duality
relation, that is $F_{5}=*F_{5}$, which in our ansatz reduces to\begin{equation}
G_{2}=*_{4}\tilde{G}_{2}.\label{eq:selfdual-g}\end{equation}
The other is the Bianchi identity $dF_{5}=0$. 

Supersymmetry requires the existence of a Killing spinor $\eta$ the
conditions for which are the Killing spinor equation \begin{equation}
\nabla_{M}\eta+\frac{i}{480}F_{_{M_{1}M_{2}M_{3}M_{4}M_{5}}}\Gamma^{M_{1}M_{2}M_{3}M_{4}M_{5}}\Gamma_{M}\eta=0\label{eq:killing-spinor-eq}\end{equation}
and the chirality condition $\Gamma_{11}\eta=\eta$ where $\Gamma_{11}\equiv\Gamma^{\bar{0}}\cdots\Gamma^{\bar{9}}$.
To analyze these conditions, we decompose the Dirac matrices in ten
dimensions as follows.\begin{equation}
\Gamma^{\bar{\mu}}=\gamma^{\bar{\mu}}\otimes1\otimes1\otimes1\otimes1,\quad\Gamma^{\bar{\hat{a}}}=\gamma_{5}\otimes\tau_{\hat{a}}\otimes1\otimes\hat{\tau}_{1},\quad\Gamma^{\bar{\tilde{a}}}=\gamma_{5}\otimes1\otimes\tau_{\tilde{a}}\otimes\hat{\tau}_{2}.\label{eq:decompose-dirac-mat}\end{equation}
Here $\gamma^{\bar{\mu}}$s are the Dirac matrices in four dimensions,
the chirality matrix in this subspace is defined as $\gamma_{5}=-i\gamma^{0}\gamma^{1}\gamma^{2}\gamma^{3}$
and $\tau_{\hat{a}},\tau_{\tilde{a}},\hat{\tau}_{1,2}$ are Pauli
matrices. We consider Killing spinors of the correspondingly decomposed
form\[
\eta=\epsilon\otimes\hat{\chi}\otimes\tilde{\chi}\]
where $\epsilon$ is a eight-component spinor on which the first and
last components of each product in \eqref{eq:decompose-dirac-mat}
act. This decomposition reduces the chirality condition for $\eta$
to\begin{equation}
\gamma_{5}\hat{\tau}_{3}\epsilon=\epsilon.\label{eq:chirality-cond}\end{equation}
Moreover we restrict $\hat{\chi}$ to a constant eigenvector of $\tau_{\hat{3}}$
and $\tilde{\chi}$ to a Killing spinor on the $S^{3}$:\begin{eqnarray*}
\tau_{\hat{3}}\hat{\chi}=s\hat{\chi} & , & s=\pm1\\
\nabla_{\bar{\tilde{a}}}'\tilde{\chi}=\frac{i}{2}b\tau_{\tilde{a}}\tilde{\chi} & , & b=\pm1\end{eqnarray*}
where $\nabla'_{\tilde{a}}$s are the covariant derivatives in the
unit radius $S^{3}$. From this point we use $\mu,\nu,\cdots$ to
denote tensors with their indices raised or lowered by the metric
of four-dimensional subspace $g_{\mu\nu}$. The Killing spinor equation
\eqref{eq:killing-spinor-eq} is expressed as follows. \begin{align}
 & \left[\nabla'_{\rho}-\frac{1}{4}s\rho_{3}F_{\rho\nu}\gamma^{\nu}\gamma_{5}^{}\hat{\tau}_{1}+isA_{\rho}-\left(\frac{1}{4}\tilde{G}_{\mu\nu}\gamma^{\mu\nu}+\frac{1}{2}s\tilde{V}_{\mu}\gamma^{\mu}\gamma_{5}\hat{\tau}_{1}+\frac{i}{2}s\tilde{g}\right)\gamma_{5}\hat{\tau}_{2}\gamma_{\rho}\right]\epsilon=0\label{eq:killing-spinor-eq-1}\\
 & \left[\frac{i}{2}\frac{\rho_{3}}{\rho_{1}}\gamma_{5}\hat{\tau}_{1}+\frac{1}{2}\not\partial\rho_{1}+\rho_{1}\left(\frac{1}{4}\tilde{G}_{\mu\nu}\gamma^{\mu\nu}+\frac{1}{2}s\tilde{V}_{\mu}\gamma^{\mu}\gamma_{5}\hat{\tau}_{1}-\frac{i}{2}s\tilde{g}\right)\gamma_{5}\hat{\tau}_{2}\right]\epsilon=0\label{eq:killing-spinor-eq-2}\\
 & \left[\frac{i}{2}\left(2-\frac{\rho_{3}^{2}}{\rho_{1}^{2}}\right)\gamma_{5}\hat{\tau}_{1}+\frac{1}{2}\not\partial\rho_{3}+\frac{1}{8}s\rho_{3}^{2}F_{\mu\nu}\gamma^{\mu\nu}\gamma_{5}\hat{\tau}_{1}+\rho_{3}\left(\frac{1}{4}\tilde{G}_{\mu\nu}\gamma^{\mu\nu}-\frac{1}{2}s\tilde{V}_{\mu}\gamma^{\mu}\gamma_{5}\hat{\ \tau}_{1}+\frac{i}{2}s\tilde{g}\right)\gamma_{5}\hat{\tau}_{2}\right]\nonumber \\
 & \qquad\qquad\qquad\qquad\qquad\qquad\qquad\qquad\qquad\qquad\qquad\qquad\qquad\qquad\qquad\qquad\times\epsilon=0\label{eq:killing-spinor-eq-3}\\
 & \left[\frac{i}{2}b\gamma_{5}\hat{\tau}_{2}+\frac{1}{2}\not\partial\tilde{\rho}-\tilde{\rho}\left(\frac{1}{4}\tilde{G}_{\mu\nu}\gamma^{\mu\nu}+\frac{1}{2}s\tilde{V}_{\mu}\gamma^{\mu}\gamma_{5}\hat{\tau}_{1}+\frac{i}{2}s\tilde{g}\right)\gamma_{5}\hat{\tau}_{2}\right]\epsilon=0\label{eq:killing-spinor-eq-4}\end{align}
where $\nabla'_{\mu}$s are the covariant derivatives in the four-dimensional
slice and $F_{\mu\nu}\equiv\partial_{\mu}A_{\nu}-\partial_{\nu}A_{\mu}$
(we will denote this two-form as $F_{2}$ in many other places in
this paper).

\subsubsection*{\emph{Analysis of the conditions}}

To extract constraints for the metric and five-form flux from the
conditions for supersymmetry, we use real spinor bilinears\begin{equation}
K_{\mu}=\bar{\epsilon}\gamma_{\mu}\epsilon,\quad L_{\mu}=\bar{\epsilon}\gamma_{5}\gamma_{\mu}\epsilon,\quad Y_{\mu\nu}=\bar{\epsilon}\gamma_{\mu\nu}\hat{\tau}_{1}\epsilon,\quad f_{1}=i\bar{\epsilon}\hat{\tau}_{1}\epsilon,\quad f_{2}=i\bar{\epsilon}\hat{\tau}_{2}\epsilon\label{eq:def-of-spinor-bilinears}\end{equation}
where $\bar{\epsilon}\equiv\epsilon^{\dagger}\gamma_{\bar{0}}$. Using
Fierz rearrangements, we can show that\begin{equation}
K^{2}=-L^{2}=-f_{1}^{2}-f_{2}^{2},\quad K\cdot L=0.\label{eq:alg-rel}\end{equation}

From the reduced Killing spinor equations \eqref{eq:killing-spinor-eq-1}\eqref{eq:killing-spinor-eq-2}\eqref{eq:killing-spinor-eq-3}\eqref{eq:killing-spinor-eq-4},
we can deduce various constraints for the components in \eqref{eq:ansatz-met}\eqref{eq:ansatz-flux}.
One of them is\[
L=-\frac{\rho_{1}f_{1}}{\rho_{3}\tilde{\rho}}dy\]
where $y\equiv\rho_{1}\tilde{\rho}$. Thus, regarding $y$ as a coordinate,
we see that $L_{y}$ is the only non-vanishing component of $L$.
Another constraint is\begin{equation}
\nabla'_{\mu}K_{\nu}=-G_{\mu\nu}f_{2}+\tilde{G}_{\mu\nu}f_{1}-\frac{\rho_{3}}{2}F_{\mu\nu}f_{2}s+\epsilon_{\mu\nu\rho\sigma}K^{\rho}V^{\sigma}s-\tilde{g}Y_{\mu\nu}s.\label{eq:derivative-of-k}\end{equation}
From this we see that $K^{\mu}$ is a Killing vector and hence it
is possible to introduce a coordinate $t$ such that $K^{\mu}\partial_{\mu}=\partial_{t}$.
Using the remaining two coordinate degrees of freedom, the metric
of the four-dimensional subspace which respects \eqref{eq:alg-rel}
reduces to \[
ds_{4}^{2}=-\frac{1}{h^{2}}\left(dt+V_{i}dx_{i}^{}\right)^{2}+h^{2}\frac{\rho_{1}^{2}}{\rho_{3}^{2}}\left(\delta_{\bar{i}\ \bar{j}}\ \tilde{e}_{i}^{\ \bar{i}}\tilde{e}_{j}^{\ \bar{j}}dx_{i}^{}dx_{j}^{}+dy^{2}\right)\]
where $i,j$ take values $1,2$, and $h^{-2}=f_{1}^{2}+f_{2}^{2}$.

Further investigations of \eqref{eq:killing-spinor-eq} show that
$\rho_{1},\rho_{3},\tilde{\rho}$ are $t$-independent and that all
the spinor bilinears defined in \eqref{eq:def-of-spinor-bilinears}
and all the components of the five-form flux and $F_{2}$ can be written
in terms of $\rho_{1},\rho_{3},\tilde{\rho},K_{\mu},A_{t}$ and the
Levi-Civita symbol $\epsilon_{\mu\nu\rho\sigma}$. For later convenience
we present here the results of $f_{1},f_{2},\tilde{V}$ and $F_{2}$.\begin{align}
f_{1}=\tilde{\rho} & ,\quad f_{2}=\rho_{3}\left(c+sA_{t}\right),\quad\tilde{V}=\frac{s}{4}\frac{1}{\rho_{3}\tilde{\rho}^{3}}d\left(b\rho_{1}^{2}\tilde{\rho}^{2}-\rho_{3}\tilde{\rho}^{2}f_{2}\right),\nonumber \\
F_{\mu\nu}= & -\frac{2s}{\rho_{3}(f_{1}^{2}+f_{2}^{2})}\left[\left(2-\frac{\rho_{3}^{2}}{\rho_{1}^{2}}\right)\frac{1}{\rho_{3}}\epsilon_{\mu\nu}^{\,\,\,\,\,\rho\sigma}K_{\rho}L_{\sigma}+\frac{b}{\tilde{\rho}}\left(K_{\mu}L_{\nu}-K_{\nu}L_{\mu}\right)\right.\nonumber \\
 & +f_{1}\epsilon_{\mu\nu}^{\,\,\,\,\rho\sigma}K_{\rho}\partial_{\sigma}\ln\left(\rho_{3}\tilde{\rho}\right)+f_{2}\left(K_{\mu}\partial_{\nu}\ln\left(\rho_{3}\tilde{\rho}\right)-K_{\nu}\partial_{\mu}\ln\left(\rho_{3}\tilde{\rho}\right)\right)\nonumber \\
 & \left.+2sf_{1}\left(K_{\mu}\tilde{V}_{\nu}-K_{\nu}\tilde{V_{\mu}}\right)-2sf_{2}\epsilon_{\mu\nu}^{\,\,\,\,\,\rho\sigma}K_{\rho}\tilde{V}_{\sigma}\right]\label{eq:f2-determined}\end{align}
where $c$ is an integral constant of the differential equation for
$f_{2}$. In solving the differential equation for $f_{1}$, noting
that the sign of $f_{1}$ is flipped by the redefinition $\epsilon\rightarrow\hat{\tau}_{3}\epsilon$
without the chirality condition \eqref{eq:chirality-cond} affected,
we have set $f_{1}$ positive, and in solving the differential equation
for $f_{2}$, noting that $F_{\mu\nu}$ is $t$-independent, we have
chosen a gauge in which $A_{\mu}$ is $t$-independent. We now set
$A_{y}=0$ by using the remaining gauge degrees of freedom.

Next we consider the constraints for the eight-component spinor $\epsilon$.
We have three projection conditions and hence one complex degrees
of freedom is left for $\epsilon$. The first projection is the chirality
condition \eqref{eq:chirality-cond}. The second comes from the relative
normalization of $K_{\bar{0}}$ and $L_{\bar{3}}$%
\footnote{Throughout this paper we take the vierbein of the four-dimensional
subspace as its non-vanishing components are given by\begin{align*}
e_{t}^{\ \bar{0}} & =\frac{1}{h},\quad e_{x_{i}}^{\ \bar{0}}=\frac{V_{i}}{h},\quad e_{x_{i}}^{\ \bar{j}}=h\frac{\rho_{1}}{\rho_{3}}\tilde{e}_{i}^{\ \bar{j}},\quad e_{y}^{\ \bar{3}}=h\frac{\rho_{1}}{\rho_{3}}.\end{align*}
}, and the third comes from the sum of \eqref{eq:killing-spinor-eq-2}
and \eqref{eq:killing-spinor-eq-4} divided by $\rho_{1,}\tilde{\rho}$
respectively. To express these conditions in a simple way, we use
a spinor $\epsilon_{0}\equiv f_{2}^{-1/2}e^{-i\delta\gamma_{5}\gamma_{3}\hat{\tau}_{1}}\epsilon$
where $\delta$ is defined by $\sinh2\delta=f_{1}/f_{2}$. The results
are\begin{equation}
\gamma_{5}\hat{\tau}_{3}\epsilon_{0}=\epsilon_{0},\quad\gamma_{\bar{1}}\gamma_{\bar{2}}\epsilon_{0}=-i\epsilon_{0},\quad\gamma_{\bar{3}}\hat{\tau}_{1}\epsilon_{0}=\epsilon_{0}\label{eq:projection-for-epsilon0}\end{equation}
and the normalization of $\epsilon_{0}$ is given by $\epsilon_{0}^{\dagger}\epsilon_{0}=1$.
Let us take an explicit representation of the Dirac matrices\begin{equation}
\gamma^{\bar{0}}=i\left(\begin{array}{cc}
 & 1\\
1\end{array}\right),\quad\gamma^{\bar{1}}=\left(\begin{array}{cc}
\tau_{1}\\
 & -\tau_{1}\end{array}\right),\quad\gamma^{\bar{2}}=\left(\begin{array}{cc}
\tau_{2}\\
 & -\tau_{2}\end{array}\right),\quad\gamma^{\bar{3}}=\left(\begin{array}{cc}
\tau_{3}\\
 & -\tau_{3}\end{array}\right)\label{eq:explicit-reps-of-dirac-mat}\end{equation}
where $\tau_{1,2,3}$ are Pauli matrices. Then the solution of \eqref{eq:projection-for-epsilon0}
is\begin{equation}
\epsilon_{0}\propto\left(\begin{array}{c}
0\\
1\\
0\\
i\\
\hline 0\\
-1\\
0\\
i\end{array}\right)\label{eq:constant-spinor1}\end{equation}
 where we have expressed the components of the spinor in a manner
that the Dirac matrices \eqref{eq:explicit-reps-of-dirac-mat} act
on the four elements in each block and $\hat{\tau}_{1,2,3}$ act on
the two blocks.

In addition to the bilinears defined in \eqref{eq:def-of-spinor-bilinears},
we can define another type of bilinears by transposing the spinors.
Note that\begin{equation}
\frac{i}{\sqrt{2}}\left(\hat{\tau}_{2}+\hat{\tau}_{3}\right)\epsilon_{0}\propto\frac{i-1}{\sqrt{2}}\left(\begin{array}{c}
0\\
1\\
0\\
1\\
\hline 0\\
1\\
0\\
-1\end{array}\right).\label{eq:rotation-with-pauli-mat}\end{equation}
We can remove the phase factor of this expression by a phase shift
or a local Lorentz rotation generated by $\gamma^{\bar{1}}\gamma^{\bar{2}}$.
Calling this factor $e^{i\lambda},$ we obtain a spinor $\epsilon_{0}'\equiv e^{-i\lambda}\frac{i}{\sqrt{2}}\left(\hat{\tau}_{2}+\hat{\tau}_{3}\right)\epsilon_{0}$
with the following properties.\begin{equation}
\epsilon_{0}^{'t}\epsilon'_{0}=1,\quad\gamma_{5}\hat{\tau}_{2}\epsilon_{0}'=\epsilon_{0}',\quad\gamma_{\bar{1}}\gamma_{\bar{2}}\epsilon_{0}'=-i\epsilon_{0}',\quad\gamma_{\bar{3}}\hat{\tau}_{1}\epsilon_{0}'=-\epsilon_{0}'.\label{eq:projection-for-epsilon0p}\end{equation}
 We now define non-vanishing spinor bilinears%
\footnote{The rotation by Pauli matrices in \eqref{eq:rotation-with-pauli-mat}
is important in defining $\omega_{\mu}$s. Note that the chirality
condition $\gamma_{5}\hat{\tau}_{3}\epsilon=\epsilon$ gives \[
\epsilon^{t}\gamma^{\bar{2}}\gamma_{\bar{\mu}}\epsilon=-\epsilon^{t}\gamma^{\bar{2}}\gamma_{5}\gamma_{\bar{\mu}}\hat{\tau}_{3}\epsilon.\]
In our representation of Dirac matrices \eqref{eq:explicit-reps-of-dirac-mat},
$\gamma^{\bar{2}}$ is antisymmetric and the others are symmetric
and hence $\gamma^{\bar{2}}\gamma_{5}\gamma_{\bar{\mu}}$ is anti-symmetric.
This implies that the above bilinears must vanish. In contrast, the
rotation \eqref{eq:rotation-with-pauli-mat} changes the chirality
condition to the second expression in \eqref{eq:projection-for-epsilon0p}
and for this reason we have non-vanishing components of $\omega$. %
}\[
\omega_{\mu}=\epsilon^{'t}\gamma^{\bar{2}}\gamma_{\mu}^{}\epsilon',\ \quad W_{\mu\nu}^{1}=\epsilon^{'t}\gamma^{\bar{2}}\gamma_{\mu\nu}\hat{\tau}_{1}\epsilon',\quad W_{\mu\nu}^{3}=\epsilon^{'t}\gamma^{\bar{2}}\gamma_{\mu\nu}\hat{\tau}_{3}\epsilon'\]
where $\epsilon'\equiv e^{-i\delta\gamma_{5}\gamma_{3}\hat{\tau}_{1}}f_{2}^{1/2}\epsilon_{0}^{'}$
$\left(=ie^{-i\lambda}\left(\hat{\sigma}_{2}+\hat{\sigma}_{3}\right)\epsilon/\sqrt{2}\right)$.
From the Killing spinor equation \eqref{eq:killing-spinor-eq}, we
obtain \begin{align}
\partial_{\mu}\omega_{\nu}-\partial_{\nu}\omega_{\mu} & =\frac{1}{\rho_{3}}\left(2-\frac{\rho_{3}^{2}}{\rho_{1}^{2}}\right)W_{\mu\nu}^{3}+\left(\frac{2b}{\tilde{\rho}}-\frac{\rho_{3}}{\tilde{\rho}\rho_{1}^{2}}f_{2}\right)W_{\mu\nu}^{1}\nonumber \\
 & \quad+\frac{1}{\rho_{3}\tilde{\rho}}\left[\omega_{\mu}\partial_{\nu}\left(\rho_{3}\tilde{\rho}\right)-\omega_{\nu}\partial_{\mu}\left(\rho_{3}\tilde{\rho}\right)\right]-2is\left(A_{\mu}\omega_{\nu}-A_{\nu}\omega_{\mu}\right).\label{eq:derivative-of-omega}\end{align}
The $(y,x^{i})$ component of this relation implies that\[
\partial_{y}\left(i\tilde{e}_{i}^{\ \bar{1}}+\tilde{e}_{i}^{\ \bar{2}}\right)=D\left(i\tilde{e}_{i}^{\ \bar{1}}+\tilde{e}_{i}^{\ \bar{2}}\right)\]
 where \[
D=h^{2}\left[2\frac{\rho_{1}\tilde{\rho}}{\rho_{3}^{2}}-2\frac{\tilde{\rho}}{\rho_{1}}+f_{2}\left(2\frac{b\rho_{1}}{\rho_{3}\tilde{\rho}}-2\frac{f_{2}}{\tilde{\rho}\rho_{1}}\right)\right].\]
From this it turns out that, performing a $y$-independent coordinate
transformation for $x_{1},x_{2}$, we can set the metric of $\left(x_{1},x_{2}\right)$
space proportional to $\delta_{ij}$. Therefore the metric of the
four-dimensional subspace reduces to\begin{equation}
ds_{4}^{2}=-\frac{1}{h^{2}}\left(dt+V_{i}dx_{i}^{}\right)^{2}+h^{2}\frac{\rho_{1}^{2}}{\rho_{3}^{2}}\left(T^{2}(x,y)\left(dx_{1}^{2}+dx_{2}^{2}\right)+dy^{2}\right)\label{eq:four-dimensional-metric}\end{equation}
 where $T$ satisfies a differential equation\begin{equation}
\partial_{y}\ln T=D.\label{eq:T-D}\end{equation}
 The $(x^{1},x^{2})$ component of \eqref{eq:derivative-of-omega}
implies that\begin{equation}
sA_{i}=\left(sA_{t}+c-b\right)V_{i}+\frac{1}{2}\epsilon_{ij}\partial_{j}\ln T\label{eq:a_i-from-omega}\end{equation}
 and the $\left(t,x_{i}^{}\right)$ components of \eqref{eq:derivative-of-omega}
imply that $c=b$. 

Assembling the above results, we can write the components of the metric
and five-form flux in a concise way. To do that we introduce three
functions $m,n,p$ which are defined by\begin{align}
\rho_{1}^{4} & =\frac{mp+n^{2}}{m}y^{4},\quad\rho_{3}^{4}=\frac{p^{2}}{m(mp+n^{2)}},\quad A_{t}=bs\frac{n-p}{p}.\label{eq:def-of-mnp}\end{align}
We can see that all the components in \eqref{eq:ansatz-met}\eqref{eq:ansatz-flux}
are expressed in terms of $m,n,p$ and $T$. The easiest to see is\begin{equation}
D=2y\left(n+m-\frac{1}{y^{2}}\right).\label{eq:d-mnpt0}\end{equation}
Eq.\eqref{eq:derivative-of-k} determines the metric component $V$
as follows.\begin{equation}
dV=by*_{3}\left[dn+\left(nD+2ym(n-p)+\frac{2n}{y}\right)dy\right].\label{eq:dv-mnpt0}\end{equation}
Here $*_{3}$ is the Hodge dual in the three-dimensional subspaces
spanned by $x_{1},x_{2}$ and $y$, the metric for which is given
by the expressions inside the bracket of the second term in \eqref{eq:four-dimensional-metric}.
From the Bianchi identities for $F_{5}$, it turns out that the following
two forms are closed and hence it is possible to define the potentials
for them.\begin{eqnarray*}
\rho_{1}^{2}\rho_{3}G_{2} & = & d(B_{t}(dt+V)+\hat{B})\\
\tilde{\rho}^{3}\tilde{G}_{2}+\frac{1}{2}\tilde{g}\rho_{1}^{2}\tilde{\rho}^{3}F_{2} & = & d(\tilde{B_{t}}(dt+V)+\hat{\tilde{B}}).\end{eqnarray*}
As we mentioned above \eqref{eq:f2-determined}, the fluxes are expressed
by other degrees of freedom. Using those expressions, we obtain \begin{align}
B_{t} & =b\frac{y^{2}}{4}\frac{n}{m},\quad d\hat{B}=\frac{y^{3}}{4}*_{3}\left[dp+4yn(p-n)dy\right]\nonumber \\
\tilde{B_{t}} & =\frac{y^{2}}{4}\frac{n-\frac{1}{y^{2}}}{p},\quad d\hat{\tilde{B}}=b\frac{y^{3}}{4}*_{3}\left[dm+2mDdy\right].\label{eq:result-for-flux}\end{align}
Thus we have succeeded in writing all the components of the metric
and five-form flux in terms of the four function $m,n,p,T$. 

Actually there are constraints other than the one that the geometry
is expressed by $m,n,p$ and $T$ in the above way. One is \eqref{eq:T-D}.
We can find constraints also from the integrability of the expressions
for $dV,d\hat{B}$ and $d\hat{\tilde{B}}$ \begin{eqnarray}
ddV & = & 0\label{eq:ddV}\\
dd\hat{B} & = & 0\label{eq:ddhB}\\
dd\hat{\tilde{B}} & = & 0.\label{eq:ddhtB}\end{eqnarray}
Explicit forms of these differential equations are given later in
this section.

The analysis to this point is essentially included in \cite{Gava:2006pu}.
Since we have four differential equations for four functions $m,n,p,T$,
we see that the whole dependence of the metric and the flux on the
coordinates is determined (at least locally) by the boundary conditions
for these functions on a plane in the $x_{1}x_{2}y$ space, which
is a generalization of the result in \cite{Lin:2004nb} where we had
one function and one differential equation imposed on it.

\subsubsection*{\emph{New constraint}}

However, we point out here that an additional constraint must be imposed
on $m,n,p$ and $T$ so that the allowed solutions are more restricted.
Note that \eqref{eq:a_i-from-omega} determines $A_{i}$ with respect
to $m,n,p$ and $T$, and recall that we have obtained \eqref{eq:f2-determined}
before and that equation determines the field strength $F_{2}\equiv dA$
in terms of $m,n,p$ and $T$. Explicitly, from \eqref{eq:f2-determined}
we obtain\begin{equation}
F=-bs\left(dt+V\right)\wedge d\left(\frac{n}{p}\right)-\frac{s}{2}*_{3}\left[\left(4m-\frac{\left(n^{2}+mp\right)\left(4n+8m\right)}{p}y^{2}\right)dy-\frac{2n}{p}ydn-2ydm\right].\label{eq:f-mnpt}\end{equation}
This must coincide with the expressions for $F_{yi},F_{ij}$ obtained
by differentiating \eqref{eq:a_i-from-omega}, that is\begin{align*}
F_{yi} & =\partial_{y}\left(A_{t}V_{i}\right)+\frac{s}{2}\epsilon_{ij}\partial_{j}\partial_{y}\ln T\\
 & =bs\partial_{y}\left(\frac{n}{p}\right)V_{i}+s\frac{n-p}{p}y\epsilon_{ij}\partial_{j}n+\frac{s}{2}\epsilon_{ij}\partial_{j}\left[2y\left(n+m-\frac{1}{y^{2}}\right)\right]\\
F_{12} & =\partial_{1}\left(A_{t}V_{2}\right)-\partial_{2}\left(A_{t}V_{1}\right)+\frac{s}{2}\left(-\partial_{1}^{2}-\partial_{2}^{2}\right)\ln T\\
 & =bs\partial_{1}\left(\frac{n}{p}\right)V_{2}-bs\partial_{2}\left(\frac{n}{p}\right)V_{1}+s\frac{n-p}{p}yT^{2}\left(nD+2ym(n-p)+\frac{2n}{y}\right)-\frac{s}{2}\left(\partial_{1}^{2}+\partial_{2}^{2}\right)\ln T\end{align*}
where we have used \eqref{eq:def-of-mnp}, \eqref{eq:d-mnpt0} and
\eqref{eq:dv-mnpt0}. The comparison for $F_{yi}$ gives no new information.
The comparison for $F_{ij}$ gives a new constraint\begin{equation}
\frac{1}{2}\left(\partial_{1}^{2}+\partial_{2}^{2}\right)\ln T=-T^{2}y\partial_{y}n-T^{2}y\partial_{y}m+2T^{2}\left(m-2m^{2}y^{2}-4mny^{2}-n^{2}y^{2}+mpy^{2}\right).\label{eq:new-constraint}\end{equation}
One might suspect that \eqref{eq:new-constraint} can be derived from
\eqref{eq:T-D}\eqref{eq:ddV}\eqref{eq:ddhB}\eqref{eq:ddhtB} and
is not a new constraint. In Section \ref{sec:s3xs3}, we will exclude
this possibility by presenting a solution for \eqref{eq:T-D}\eqref{eq:ddV}\eqref{eq:ddhB}\eqref{eq:ddhtB}
which does not solve \eqref{eq:new-constraint} (see below \eqref{eq:reduced-consistency5}).

\subsubsection*{\emph{Summary}}

Here we summarize the result of this section. In the remainder of
the paper, we set $b=s=1$. The expressions for the metric and the
five-form flux are\begin{eqnarray}
ds^{2} & = & -h^{-2}\left(dt+V_{i}dx_{i}\right)+h^{2}\frac{\rho_{_{1}}^{2}}{\rho_{3}^{2}}\left(T^{2}\left(dx_{1}^{2}+dx_{2}^{2}\right)+dy^{2}\right)+\tilde{\rho}^{2}d\tilde{\Omega}_{3}^{2}\label{eq:gmno-met}\\
 &  & +\rho_{1}^{2}\left(\hat{\sigma}_{1}^{2}+\hat{\sigma}_{2}^{2}\right)+\rho_{3}^{2}\left(\hat{\sigma}_{3}-A_{t}dt-A_{i}dx^{i}\right)^{2}\nonumber \\
F_{5} & = & -\left(G_{mn}e^{m}\wedge e^{n}\wedge e^{\hat{1}}\wedge e^{\hat{2}}\wedge e^{\hat{3}}+*_{4}\tilde{V}\wedge e^{\hat{1}}\wedge e^{\hat{2}}+*_{4}\tilde{g}\wedge e^{\hat{3}}\right)\label{eq:gmno-flux}\\
 &  & +\left(\tilde{G}_{mn}e^{m}\wedge e^{n}+\tilde{V}_{m}e^{m}\wedge e^{\hat{3}}+\tilde{g}e^{\hat{1}}\wedge e^{\hat{2}}\right)\wedge\tilde{\rho}^{3}d\tilde{\Omega}_{3}.\nonumber \end{eqnarray}
$h^{2}$ and the components of the five-form are expressed by $\rho_{1},\rho_{3},\tilde{\rho},V,A$
(or its field strength $F_{2}\equiv dA$),$B_{t},\hat{B},\tilde{B}_{t}$
and $\hat{\tilde{B}}$. \begin{eqnarray}
h^{-2} & = & \tilde{\rho}^{2}+\rho_{3}^{2}\left(1+A_{t}\right)^{2}\nonumber \\
\tilde{g} & = & \frac{1}{2\tilde{\rho}}\left(1-\frac{\rho_{3}^{2}}{\rho_{1}^{2}}\left(1+A_{t}\right)\right)\nonumber \\
\tilde{V} & = & \frac{1}{2\rho_{3}\tilde{\rho}^{3}}d\left(\tilde{g}\rho_{1}^{2}\tilde{\rho}^{3}\right)\nonumber \\
\rho_{1}^{2}\rho_{3}G & = & d(B_{t}(dt+V)+\hat{B})\nonumber \\
\tilde{G}\tilde{\rho}^{3} & = & -\frac{1}{2}\tilde{g}\rho_{1}^{2}\tilde{\rho}^{3}F_{2}+d(\tilde{B_{t}}(dt+V)+\hat{\tilde{B}}).\label{eq:def-of-bttbth}\end{eqnarray}
The remaining degrees of freedom are further reduced to $m,n,p$ and
$T$ by the following relations.\begin{eqnarray}
\rho_{1}^{4}=\frac{mp+n^{2}}{m}y^{4} & , & \rho_{3}^{4}=\frac{p^{2}}{m(mp+n^{2)}}\nonumber \\
\tilde{\rho}^{4}=\frac{m}{mp+n^{2}} & , & A_{t}=\frac{n-p}{p}\label{eq:rhotat-mtnp}\end{eqnarray}
\begin{eqnarray}
dV & = & y*_{3}\left[dn+\left(nD+2ym(n-p)+\frac{2n}{y}\right)dy\right]\label{eq:dV-mnpt}\\
A_{i} & = & A_{t}V_{i}+\frac{1}{2}\epsilon_{ij}\partial_{j}\ln T\label{eq:ai-mnpt}\end{eqnarray}
\begin{eqnarray}
B_{t}=\frac{y^{2}}{4}\frac{n}{m} & , & d\hat{B}=\frac{y^{3}}{4}*_{3}\left[dp+4yn(p-n)dy\right]\label{eq:btdbh-mnpt}\\
\tilde{B_{t}}=\frac{y^{2}}{4}\frac{n-\frac{1}{y^{2}}}{p} & , & d\hat{\tilde{B}}=\frac{y^{3}}{4}*_{3}\left[dm+2mDdy\right],\label{eq:bttdbht-mnpt}\end{eqnarray}
 where $D=2y(m+n-1/y^{2})$. We have five differential equations for
$m,n,p,T$\begin{align}
 & y^{3}\left(\partial_{1}^{2}+\partial_{2}^{2}\right)n+\partial_{y}\left(y^{3}T^{2}\partial_{y}n\right)+y^{2}\partial_{y}\left[T^{2}\left(yDn+2y^{2}m(n-p)\right)\right]+4y^{2}DT^{2}n=0\label{eq:consistency1}\\
 & y^{3}\left(\partial_{1}^{2}+\partial_{2}^{2}\right)m+\partial_{y}\left(y^{3}T^{2}\partial_{y}m\right)+\partial_{y}\left(2y^{3}T^{2}mD\right)=0\label{eq:consistency2}\\
 & y^{3}\left(\partial_{1}^{2}+\partial_{2}^{2}\right)p+\partial_{y}\left(y^{3}T^{2}\partial_{y}p\right)+\partial_{y}\left[4y^{3}T^{2}ny(n-p)\right]=0\label{eq:consistency3}\\
 & \partial_{y}\ln T=D.\label{eq:consistency4}\end{align}
\begin{equation}
\frac{1}{2}\left(\partial_{1}^{2}+\partial_{2}^{2}\right)\ln T=-T^{2}y\partial_{y}n-T^{2}y\partial_{y}m+2T^{2}\left(m-2m^{2}y^{2}-4mny^{2}-n^{2}y^{2}+mpy^{2}\right).\label{eq:consistency5}\end{equation}
(\eqref{eq:consistency1}, \eqref{eq:consistency2} and \eqref{eq:consistency3}
are the explicit forms of \eqref{eq:ddV}, \eqref{eq:ddhB} and \eqref{eq:ddhtB}
respectively.)

We have written down many constraints derived from the Bianchi identity,
the self-duality relation and the Killing spinor equation, but it
is uncertain whether we have equivalently transformed those original
constraints. Moreover we need to impose the Einstein equation\begin{equation}
R_{\mu\nu}=\frac{1}{6}F_{\mu\alpha\beta\gamma\delta}F_{\ \quad\ \ \nu}^{\alpha\beta\gamma\delta}\label{eq:einstein-eq}\end{equation}
on the above geometries. In the next section, we work on this issue
for a restricted case of $m$ and $n$, and show that the above results
are insufficient to produce solutions of the supergravity with the
symmetries required in the setup.

\section{Deviation from LLM with $D=0,\ \rho_{1}=\rho_{3},$ and $n$ fixed\label{sec:s3xs3}}

The result in the previous section is a generalization of that in
\cite{Lin:2004nb}(LLM). A limit to LLM solutions is given by $\rho_{3}=\rho_{1},A_{t}=0,T=const.$,
in other words it is $D=0,n=p,T=const$.. In this limit, all the degrees
of freedom reduce to one function and the differential equation imposed
on it can be solved by integral forms for general boundary conditions.
However, in general case, although we have obtained differential equations
\eqref{eq:consistency1}\eqref{eq:consistency2}\eqref{eq:consistency3}\eqref{eq:consistency4}\eqref{eq:consistency5}
for the controlling functions $m,n,p,T$, it is far more difficult
to solve them or find physical implications from them. Therefore we
seek limits in which those equations reduce to tractable forms such
that we can find meaningful information from them. 

One of the chief interests would be on the property of our geometries
near LLM ansatz. Paying attention to \eqref{eq:rhotat-mtnp}, we find
that setting $\rho_{1}=\rho_{3}$ almost gives another $S^{3}$ metric
but this condition leaves two of the three degrees of freedom $m,n,p$.
If we further set $D=0$, $n-p$ is left as a deformation function
for a special case of LLM. Expanding \eqref{eq:rhotat-mtnp} in $n-p$
to the first order, we obtain\begin{eqnarray*}
\rho_{1}^{4} & \sim & \frac{ny^{4}}{1-y^{2}n}-y^{4}(n-p)\\
\rho_{3}^{4} & \sim & \frac{ny^{4}}{1-y^{2}n}-\frac{1+ny^{2}}{1-ny^{2}}y^{4}(n-p).\end{eqnarray*}
 Equating these two gives $n=0,m=1/y^{2}$. This condition is sufficient
to satisfy $\rho_{1}=\rho_{3}$ to all order and therefore we concentrate
on these continuously deviated LLM geometries which have only two
degrees of freedom $p,T$. %
\footnote{For these geometries, $n$ is fixed to 0 and $p$ deviates from the
LLM limit $n=p=0$, but there is another solution for $\rho_{1}=\rho_{3},D=0$,
in which $n$ also deviates from $0$ and $p,n$ satisfy a constraint
$p-n=2n/(y^{2}n-2)$ . In this paper, we do not investigate this case
and leave it for a future work. %
}

In this case, the differential equations \eqref{eq:consistency1}\eqref{eq:consistency2}\eqref{eq:consistency3}\eqref{eq:consistency4}\eqref{eq:consistency5}
are reduced to simple forms. Eq.\eqref{eq:consistency2} and \eqref{eq:consistency4}
are equivalent and both imply that $T$ is $y$-independent, $T=T(x)$.
Then \eqref{eq:consistency1} implies that $p$ is also $y$-independent,
$p=p(x)$. and hence \eqref{eq:consistency3} reduces to a Laplace
equation in two dimensions\begin{equation}
\left(\partial_{1}^{2}+\partial_{2}^{2}\right)p(x)=0.\label{eq:reduced-consistensy4}\end{equation}
Eq.\eqref{eq:consistency5} reduces to a simple but nonlinear equation\begin{equation}
\left(\partial_{1}^{2}+\partial_{2}^{2}\right)\ln\left(T(x)^{2}\right)=8p(x)T(x)^{2}.\label{eq:reduced-consistency5}\end{equation}
Now it is clear that \eqref{eq:consistency5} is independent from
\eqref{eq:consistency1}\eqref{eq:consistency2}\eqref{eq:consistency3}\eqref{eq:consistency4}.
In our restricted case, the constraints of \eqref{eq:consistency1}\eqref{eq:consistency2}\eqref{eq:consistency3}\eqref{eq:reduced-consistensy4}
are equivalent to the requirement that $p$ and $T$ are $y$-independent
and $ $$p$ satisfies \eqref{eq:reduced-consistensy4}, and hence
they allow $p$ and $T$ to be constant, but that does not satisfy
\eqref{eq:reduced-consistency5}. Thus we can say that \eqref{eq:consistency5}
is independent from \eqref{eq:consistency1}\eqref{eq:consistency2}\eqref{eq:consistency3}\eqref{eq:consistency4}.
Eq.\eqref{eq:reduced-consistency5} (in other words \eqref{eq:consistency5})
plays important roles in the remainder of this paper. 

The other expressions in the result of the previous section also reduced
to simple forms. We present some of them first. \eqref{eq:dV-mnpt}
becomes\begin{equation}
dV=-2pT^{2}dx^{1}\wedge dx^{2}.\label{eq:reduced-dv}\end{equation}
This equation for $V$ can be solved by using \eqref{eq:reduced-consistency5}.
The solutions are given by\begin{equation}
V_{i}=\frac{1}{4}\epsilon_{ij}\partial_{j}\ln T^{2}+\partial_{i}\alpha\label{eq:sol-for-v}\end{equation}
where the first term is a particular solution guaranteed by \eqref{eq:reduced-consistency5}
and $\alpha$ is an arbitrary function depending on $x_{1},x_{2}$.
The last expression in \eqref{eq:rhotat-mtnp} reduces to $A_{t}=-1$,
\eqref{eq:ai-mnpt} reduces to $A_{i}=-\partial_{i}\alpha$, and hence
$F_{2}=0$. 

Straightforwardly we obtain reduced expressions for the metric and
five-form flux\begin{align}
ds^{2} & =-\frac{1}{\sqrt{p}}\left(dt+V\right)^{2}+\sqrt{p}\left(T^{2}\left(x_{1}^{2}+x_{2}^{2}\right)+dy^{2}\right)+\frac{1}{\sqrt{p}}d\tilde{\Omega}_{3}^{2}\nonumber \\
 & \quad+\sqrt{p}y^{2}\left[\sigma_{\hat{1}}^{2}+\sigma_{\hat{2}}^{2}+\left(\sigma_{\hat{3}}+dt+\partial_{i}\alpha dx_{i}^{}\right)^{2}\right]\label{eq:reduced-gmno-met}\\
F_{5} & =-\rho_{1}^{2}\rho_{3}G_{2}\wedge\sigma_{\hat{1}}\wedge\sigma_{\hat{2}}\wedge\left(\sigma_{\hat{3}}+dt+\partial_{i}\alpha dx_{i}\right)\nonumber \\
 & \quad-\frac{p}{2}y^{2}*_{4}dy\wedge\sigma_{\hat{1}}\wedge\sigma_{\hat{2}}-\frac{\sqrt{p}}{2}y*_{4}1\wedge\left(\sigma_{\hat{3}}+dt+\partial_{i}\alpha dx_{i}\right)\nonumber \\
 & \quad+\left(\tilde{\rho}^{3}\tilde{G}_{2}+\frac{y}{2}dy\wedge\left(\sigma_{\hat{3}}+dt+\partial_{i}\alpha dx_{i}\right)+\frac{y^{2}}{2}\sigma_{\hat{1}}\wedge\sigma_{\hat{2}}\right)\wedge d\tilde{\Omega}_{3}\label{eq:reduced-gmno-flux}\end{align}
where\begin{eqnarray}
\rho_{1}^{2}\rho_{3}G_{2} & = & \frac{y^{3}}{4}*_{3}dp\label{eq:reduced-g}\\
\tilde{\rho}^{3}\tilde{G}_{2} & = & \frac{1}{4p^{2}}dp\wedge(dt+V).\label{eq:reduced-gt}\end{eqnarray}
At this stage we can see that the self-duality relation \eqref{eq:selfdual-g}
is restored by using the expressions \eqref{eq:reduced-g}\eqref{eq:reduced-gt}.
Note that it is due to \eqref{eq:reduced-consistency5} that we deduced
that $F_{2}=0$ and hence have the vanishing first term in \eqref{eq:def-of-bttbth}.

\subsubsection*{\emph{Complete set of constraints}}

We have written down the reduced forms of the expressions in the summary
of the previous section (\eqref{eq:gmno-met} -- \eqref{eq:consistency5}).
We now exhaust all the other constraints for the above geometries. 

First we reexamine the Killing spinor equation. In the previous section,
the form of the spinor $\epsilon$ has been partly determined. Explicitly
it is \begin{align}
\epsilon & =f_{2}^{\frac{1}{2}}e^{i\delta\gamma_{5}\gamma_{3}\hat{\tau}_{1}}\epsilon_{0}\nonumber \\
 & =e^{i\left(\lambda-\frac{3}{4}\pi\right)}f_{2}^{\frac{1}{2}}e^{i\delta\gamma_{5}\gamma_{3}\hat{\tau}_{1}}\epsilon_{c}\nonumber \\
 & =e^{i\left(\lambda-\frac{3}{4}\pi\right)}f^{\frac{1}{2}}\left(\cosh\delta+i\sinh\delta\ \hat{\tau}_{3}\right)\epsilon_{c}\label{eq:partially-determined-epsilon}\end{align}
where $\epsilon_{c}$ is the constant spinor in the right hand side
of \eqref{eq:constant-spinor1}. In the third line we have used projection
conditions in \eqref{eq:projection-for-epsilon0}. From the expressions
for $f_{1},f_{2}$ in \eqref{eq:f2-determined}, we see that in our
limit $f_{2}$ vanishes, hence $e^{\delta}$ diverges as $e^{\delta}\sim2\left(\frac{f_{1}}{f_{2}}\right)^{1/2}$
and \eqref{eq:partially-determined-epsilon} converges to\[
\epsilon\sim e^{i\left(\lambda-\frac{3}{4}\pi\right)}f_{1}^{1/2}\left(1+i\hat{\tau}_{3}\right)\epsilon_{c}=p^{-1/8}e^{i\left(\lambda-\frac{3}{4}\pi\right)}\left(1+i\hat{\tau}_{3}\right)\epsilon_{c}.\]
We substitute this into \eqref{eq:killing-spinor-eq-1}\eqref{eq:killing-spinor-eq-2}\eqref{eq:killing-spinor-eq-3}\eqref{eq:killing-spinor-eq-4}.
Using \eqref{eq:projection-for-epsilon0} again, we obtain \begin{align*}
\left(i\not\tilde{V}\hat{\tau}_{3}-i\tilde{g}\gamma_{5}\hat{\tau}_{2}\right)\epsilon & =\frac{1}{2}p^{\frac{1}{4}}\left(i\gamma_{\bar{3}}\hat{\tau}_{3}+\hat{\tau}_{1}\right)\epsilon\\
 & =\frac{1}{2}p^{\frac{1}{8}}e^{i\left(\lambda-\frac{3}{4}\pi\right)}\left(i\gamma_{\bar{3}}\hat{\tau}_{3}+\hat{\tau}_{1}\right)\left(1+i\hat{\tau}_{3}\right)\epsilon_{c}\\
 & =0.\end{align*}
Thus we see that \eqref{eq:killing-spinor-eq-2} and \eqref{eq:killing-spinor-eq-3}
are equivalent in our limit. Recall that the sum of \eqref{eq:killing-spinor-eq-2}
and \eqref{eq:killing-spinor-eq-4} divided by $\rho_{1},\tilde{\rho}$
is solved by the projection conditions for $\epsilon_{0}$ \eqref{eq:projection-for-epsilon0}.
Therefore we consider only \eqref{eq:killing-spinor-eq-1} and \eqref{eq:killing-spinor-eq-2}.
To reexamine \eqref{eq:killing-spinor-eq-1}, we need the expression
for the the spin connection $\omega_{\mu\bar{\nu}\bar{\rho}}$ in
the four-dimensional subspace. Its non-vanishing components are shown
to be\begin{align*}
 & \omega_{t\bar{0}\bar{1}}=-\omega_{t\bar{1}\bar{0}}=\frac{\partial_{1}p}{4p^{\frac{3}{2}}T},\quad\omega_{t\bar{0}\bar{2}}=-\omega_{t\bar{2}\bar{0}}=\frac{\partial_{2}p}{4p^{\frac{3}{2}}T},\quad\omega_{t\bar{1}\bar{2}}=-\omega_{t\bar{2}\bar{1}}=-1\\
 & \omega_{x_{1}\bar{0}\bar{1}}=-\omega_{x_{1}\bar{1}\bar{0}}=\frac{\partial_{1}p}{4p^{3/2}T}V_{1},\quad\omega_{x_{1}\bar{0}\bar{2}}=-\omega_{x_{1}\bar{2}\bar{0}}=\frac{\partial_{2}p}{4p^{3/2}T}V_{1}-p^{1/2}T\\
 & \omega_{x_{2}\bar{0}\bar{1}}=-\omega_{x_{2}\bar{1}\bar{0}}=\frac{\partial_{1}p}{4p^{3/2}T}V_{2}+p^{1/2}T,\quad\omega_{x_{2}\bar{0}\bar{2}}=-\omega_{x_{2}\bar{2}\bar{0}}=\frac{\partial_{2}p}{4p^{3/2}T}V_{2}\\
 & \omega_{x_{1}\bar{1}\bar{2}}=-\omega_{x_{1}\bar{2}\bar{1}}=\frac{\partial_{2}p}{4p}+V_{1}-2\partial_{1}\alpha,\quad\omega_{x_{2}\bar{1}\bar{2}}=-\omega_{x_{2}\bar{2}\bar{1}}=-\frac{\partial_{1}p}{4p}+V_{2}-2\partial_{2}\alpha\\
 & \omega_{y\bar{1}\bar{3}}=-\omega_{y\bar{3}\bar{1}}=-\frac{\partial_{1}p}{4pT},\quad\omega_{y\bar{2}\bar{3}}=-\omega_{y\bar{3}\bar{2}}=-\frac{\partial_{2}p}{4pT}.\end{align*}
Using these expressions and the projection conditions \eqref{eq:projection-for-epsilon0p},
we obtain the reduced forms of \eqref{eq:killing-spinor-eq-1}\begin{align*}
\partial_{t} & \left(p^{-1/8}e^{i\left(\lambda-\frac{3}{4}\pi\right)}\left(1+i\hat{\tau}_{3}\right)\epsilon_{c}\right)=0\\
\left[\partial_{x_{1}}+\frac{1}{8}\left(\partial_{x_{1}}p\right)\right] & \left(p^{-1/8}e^{i\left(\lambda-\frac{3}{4}\pi\right)}\left(1+i\hat{\tau}_{3}\right)\epsilon_{c}\right)=0\\
\left[\partial_{x_{2}}+\frac{1}{8}\left(\partial_{x_{2}}p\right)\right] & \left(p^{-1/8}e^{i\left(\lambda-\frac{3}{4}\pi\right)}\left(1+i\hat{\tau}_{3}\right)\epsilon_{c}\right)=0\\
\partial_{y} & \left(p^{-1/8}e^{i\left(\lambda-\frac{3}{4}\pi\right)}\left(1+i\hat{\tau}_{3}\right)\epsilon_{c}\right)=0,\end{align*}
which leads to that $\lambda=const.$. We can show that \eqref{eq:killing-spinor-eq-2}
reduces to a trivial equation and gives no new constraint. Thus we
see that the Killing spinor equation \eqref{eq:killing-spinor-eq}
only determines the phase factors of the Killing spinors and gives
no new constraint for the metric and five-form flux \eqref{eq:sol-for-v}\eqref{eq:reduced-gmno-met}\eqref{eq:reduced-gmno-flux}\eqref{eq:reduced-g}\eqref{eq:reduced-gt}.

Next we consider the Einstein equation \eqref{eq:einstein-eq}. For
convenience, we rewrite the expressions for the metric and five form
flux in the following way. First we perform coordinate transformations
$t\rightarrow t-\alpha,\hat{\psi}\rightarrow\hat{\psi}-t$ where $\hat{\psi}$
is a coordinate of the squashed three-sphere (see \eqref{eq:explicit-forms-of-squashed-s3}).
Note that the second transformation just absorbs the $dt$ accompanied
by $\sigma_{\hat{3}}$ and does not affect the other components of
the metric and five-form flux: \[
\sigma_{\hat{3}}^{}+dt\rightarrow\sigma_{\hat{3}}^{},\quad\left(\sigma_{\hat{1}}^{}\right)^{2}+\left(\sigma_{\hat{2}}^{}\right)^{2}\rightarrow\left(\sigma_{\hat{1}}^{}\right)^{2}+\left(\sigma_{\hat{2}}^{}\right)^{2},\quad\sigma_{\hat{1}}\wedge\sigma_{\hat{2}}\rightarrow\sigma_{\hat{1}}\wedge\sigma_{\hat{2}}.\]
We now see that another $S^{3}$ metric $d\hat{\Omega}^{2}\equiv\sigma_{\hat{1}}^{2}+\sigma_{\hat{2}}^{2}+\sigma_{\hat{3}}^{2}$
$ $appears in the metric \eqref{eq:reduced-gmno-met}. We parametrize
that $S^{3}$ with a unit vector in four-dimensional space $\hat{\bm{y}}=\left(\hat{y}_{1},\hat{y}_{2},\hat{y}_{3},\hat{y}_{4}\right)$,
regard $y$ as the coordinate of the radial direction and introduce
coordinates $y_{1,2,3,4}\equiv y\hat{y}_{1,2,3,4}$.$ $ We have the
relations\begin{align*}
dy^{2}+y^{2}\left(\sigma_{\hat{1}}^{2}+\sigma_{\hat{2}}^{2}+\sigma_{\hat{3}}^{2}\right) & =dy_{1}^{2}+dy_{2}^{2}+dy_{3}^{2}+dy_{4}^{2}\\
ydy\wedge\sigma_{\hat{3}}+y^{2}\sigma_{\hat{1}}\wedge\sigma_{\hat{2}} & =ydy\wedge\sigma_{\hat{3}}+\frac{y^{2}}{2}d\sigma_{\hat{3}}\\
 & =-\frac{1}{2}R_{\alpha\beta}^{1}\left(\hat{y}_{\alpha}^{}dy+yd\hat{y}_{\alpha}^{}\right)\wedge\left(\hat{y}_{\beta}^{}dy+yd\hat{y}_{\beta}^{}\right)\\
 & =-dy_{1}\wedge dy_{2}-dy_{3}\wedge dy_{4}\end{align*}
(see \eqref{eq:def-of-rs}\eqref{eq:r1-sigma3}). Using these for
\eqref{eq:reduced-gmno-met}\eqref{eq:reduced-gmno-flux}, we obtain
the following expressions for the metric and five-form flux.\begin{align*}
ds^{2} & =-\frac{1}{\sqrt{p}}\left(dt+V\right)^{2}+\sqrt{p}T^{2}\left(dx_{1}^{2}+dx_{2}^{2}\right)+\sqrt{p}\left(dy_{1}^{2}+dy_{2}^{2}+dy_{3}^{2}+dy_{4}^{2}\right)+\frac{1}{\sqrt{p}}d\tilde{\Omega}_{2}^{2}\\
V_{i} & =\frac{1}{4}\epsilon_{ij}\partial_{j}\ln T^{2}\\
F_{5} & =\frac{1}{2}\left(\partial_{2}pdx_{1}-\partial_{1}pdx_{2}\right)\wedge dy_{1}\wedge dy_{2}\wedge dy_{3}\wedge dy_{4}+\frac{pT^{2}}{2}dt\wedge dx_{1}\wedge dx_{2}\wedge\left(dy_{1}\wedge dy_{2}+dy_{3}\wedge dy_{4}\right)\\
 & +\frac{1}{2p^{2}}\left(\partial_{1}pdx_{1}+\partial_{2}pdx_{2}\right)\wedge\left(dt+V\right)\wedge d\tilde{\Omega}-\frac{1}{2}\left(dy_{1}\wedge dy_{2}+dy_{3}\wedge dy_{4}\right)\wedge d\tilde{\Omega}.\end{align*}
The $(t,t)$ component of \eqref{eq:einstein-eq} for this geometry
is calculated to be\begin{align*}
0 & =R_{tt}-\frac{1}{6}F_{t\alpha\beta\gamma\delta}F_{\ \quad\ \ t}^{\alpha\beta\gamma\delta}\\
 & =-\frac{3}{4p^{3}T^{2}}\left(\left(\partial_{1}p(x)\right)^{2}+\left(\partial_{2}p(x)\right)^{2}\right),\end{align*}
which implies that $\partial_{1}p=\partial_{2}p=0$, that is, $p$
is constant.

We have shown that the metric and the five form flux are expressed
with one constant parameter $p$ in the following way.\begin{align}
ds^{2} & =-\frac{1}{\sqrt{p}}\left(dt+V\right)^{2}+\sqrt{p}T^{2}\left(dx_{1}^{2}+dx_{2}^{2}\right)+\sqrt{p}\left(dy_{1}^{2}+dy_{2}^{2}+dy_{3}^{2}+dy_{4}^{2}\right)+\frac{1}{\sqrt{p}}d\tilde{\Omega}_{3}^{2}\label{eq:reduced-met1.5}\\
F_{5} & =\frac{p}{2}T^{2}dt\wedge dx_{1}\wedge dx_{2}\wedge\left(dy_{1}\wedge dy_{2}+dy_{3}\wedge dy_{4}\right)-\frac{1}{2}\left(dy_{1}\wedge dy_{2}+dy_{3}\wedge dy_{4}\right)\wedge d\tilde{\Omega}_{3}.\label{eq:reduced-flux1.5}\\
 & V=\frac{1}{4}\epsilon_{ij}\partial_{j}\ln T^{2}dx_{i}.\label{eq:sol-for-v2}\end{align}
Because $p$ is constant, the remaining known constraint \eqref{eq:reduced-consistency5}
is a Liouville equation with a cosmological constant $-16p$\begin{equation}
\left(\partial_{1}^{2}+\partial_{2}^{2}\right)\ln\left(T(x)^{2}\right)=8pT(x)^{2}.\label{eq:liouville-eq}\end{equation}
As we will see below, the solutions of this equation correspond to
geometries which are locally equivalent to the near horizon geometry
of intersecting D3-brane systems. This implies that all of them are
solutions of the supergravity and hence no additional constraint arises
from the other components of the Einstein equation \eqref{eq:einstein-eq}.

\subsubsection*{$\bm{AdS_{3}\times S^{3}\times R^{4}}$ }

The general solution of \eqref{eq:liouville-eq} has been known through
the study of two dimensional surface. On each connected domain in
$ $$x_{1}x_{2}$ space, it is of the form

\begin{equation}
T^{2}dud\bar{u}=\frac{1}{p}\frac{\partial\xi(u)\bar{\partial}\bar{\xi}\left(\bar{u}\right)}{\left|\xi(u)-\bar{\xi}\left(\bar{u}\right)\right|^{2}}dud\bar{u}\label{eq:general-sol}\end{equation}
where $u\equiv x^{1}+ix^{2}$ and $\xi$ is an arbitrary holomorphic
function. From the point of view of the global structure of the surface,
$u$ is the coordinate of a local patch inside the upper half plane
or its quotient by the discrete subgroup $\Gamma$ of the M$\Ddot{\mathrm{o}}$bius
group $SL(2,R)$ and $\xi$ is the local coordinate of the surface
with which the metric is expressed in the standard form $ds_{2}^{2}\propto d\xi d\bar{\xi}/\left(\mathrm{Im}\xi\right)^{2}$.
The solutions are classified by the matrices $M\in\Gamma$ which act
on $\xi(u)$ as $u$ goes around fixed points of $\Gamma$: 1)$\left|\mathrm{Tr}M\right|<2$
(elliptic), 2)$\left|\mathrm{Tr}M\right|=2$ (parabolic), 3)$\left|\mathrm{Tr}M\right|>2$
(hyperbolic). In this paper, we do not investigate the global structures
of the solutions, and in that case, it is sufficient to consider one
solution of \eqref{eq:liouville-eq} because the other solutions are
related to it by coordinate transformation at least locally.

Let us consider an example of parabolic solution $T^{2}=1/4px_{1}^{2}$
. Then, from \eqref{eq:reduced-met1.5}\eqref{eq:reduced-flux1.5}\eqref{eq:sol-for-v2},
we obtain the following metric and the five-form.\begin{align}
ds^{2} & =-\frac{1}{\sqrt{p}}\left(dt+\frac{1}{2x_{1}}dx_{2}\right)^{2}+\frac{1}{4\sqrt{p}x_{1}^{2}}\left(dx_{1}^{2}+dx_{2}^{2}\right)+\sqrt{p}\left(dy_{1}^{2}+dy_{2}^{2}+dy_{3}^{2}+dy_{4}^{2}\right)+\frac{1}{\sqrt{p}}d\tilde{\Omega}_{3}^{2}\label{eq:reduced-met2}\\
F_{5} & =\frac{1}{8x_{1}^{2}}dt\wedge dx_{1}\wedge dx_{2}\wedge\left(dy_{1}\wedge dy_{2}+dy_{3}\wedge dy_{4}\right)-\frac{1}{2}\left(dy_{1}\wedge dy_{2}+dy_{3}\wedge dy_{4}\right)\wedge d\tilde{\Omega}_{3}.\label{eq:reduced-flux2}\end{align}
The last two terms of the metric represent $R^{4}$ and $S^{3}$ respectively.
We can show that the three-dimensional space spanned by $\left(t,x_{1},x_{2}\right)$
is $AdS_{3}$. One way to do that is to show that its metric satisfies
three-dimensional Einstein equation with a negative cosmological constant.
This is sufficient to study local issues because the local structures
of three-dimensional gravity are governed by its cosmological constant.
Another way is to present explicit coordinate transformations which
lead to standard expressions for $AdS_{3}$, which will be more useful
for the studies of global issues in the future. One such transformation
is given by\begin{eqnarray}
x_{1} & = & \frac{z^{2}}{1+x^{+2}}\nonumber \\
x_{2} & = & x^{-}-z^{2}\frac{x^{+}}{1+x^{+2}}\nonumber \\
t & = & \arctan x^{+}\label{eq:coord-trans-to-poincare}\end{eqnarray}
and this leads to a Poincar$\Acute{\mathrm{e}}$ metric of $AdS_{3}$
with radius $1/p^{\frac{1}{4}}$.%
\footnote{We can show that the $AdS_{3}$ written in the global coordinate is
also covered by the coordinate system used in \eqref{eq:reduced-met2}\eqref{eq:reduced-flux2}. %
} Explicitly \eqref{eq:reduced-met2}\eqref{eq:reduced-flux2} become\begin{align}
ds^{2} & =\frac{1}{\sqrt{p}}\frac{-dx^{+}dx^{-}+dz^{2}}{z^{2}}+\sqrt{p}\left(dy_{1}^{2}+dy_{2}^{2}+dy_{3}^{2}+dy_{4}^{2}\right)+\frac{1}{\sqrt{p}}d\tilde{\Omega}_{3}^{2}\label{eq:reduced-met3}\\
F_{5} & =\frac{1}{4z^{3}}dx^{+}\wedge dx^{-}\wedge dz\wedge\left(dy_{1}\wedge dy_{2}+dy_{3}\wedge dy_{4}\right)-\frac{1}{2}\left(dy_{1}\wedge dy_{2}+dy_{3}\wedge dy_{4}\right)\wedge d\tilde{\Omega}_{3}.\label{eq:reduced-flux3}\end{align}

The above ten-dimensional space is the near horizon geometry of an
intersecting D3-brane system. To see this, let us consider a stack
of D3-branes such that all the branes extend in $ $$1+1$ directions
$w_{0},w_{1}$ and localize in four directions $z_{1,}z_{2},z_{3},z_{4}$
(overall transverse space) and the remaining world volume directions
are $y_{1},y_{2}$ or $y_{3,}y_{4}$ (relative transverse space).
This configuration is summarized in the following table.\\
\\
\begin{tabular}{|c||c|c|c|c|c|c|c|c|c|c|}
\hline 
 & $w_{0}$ & $w_{1}$ & $y_{1}$ & $y_{2}$ & $y_{3}$ & $y_{4}$ & $z_{1}$ & $z_{2}$ & $z_{3}$ & $z_{4}$\tabularnewline
\hline 
D3 & $ $$\circ$ & $\circ$ & $\circ$ & $\circ$ &  &  &  &  &  & \tabularnewline
\hline 
D3 & $\circ$ & $\circ$ &  &  & $\circ$ & $\circ$ &  &  &  & \tabularnewline
\hline
\end{tabular}\\
\\
The supergravity solution which in a sense corresponds to this configuration
is given as follows (see, for a review \cite{Youm:1997hw}).\begin{align}
ds^{2} & =H_{1}^{-\frac{1}{2}}H_{2}^{-\frac{1}{2}}\left(-dw_{0}^{2}+dw_{1}^{2}\right)+H_{1}^{-\frac{1}{2}}H_{2}^{\frac{1}{2}}\left(dy_{1}^{2}+dy_{2}^{2}\right)+H_{1}^{\frac{1}{2}}H_{2}^{-\frac{1}{2}}\left(dy_{3}^{2}+dy_{4}^{2}\right)\label{eq:intersection-right-met}\\
 & +H_{1}^{\frac{1}{2}}H_{2}^{\frac{1}{2}}\sum_{i=1}^{4}dz_{i}^{2}\nonumber \\
F_{5} & =-\frac{1}{2}dw_{0}\wedge dw_{1}\wedge dr\wedge\left(\frac{l_{1}}{r^{3}}H_{1}^{-2}dy_{1}^{}\wedge dy_{2}+\frac{l_{2}}{r^{3}}H_{2}^{-2}dy_{3}^{}\wedge dy_{4}\right)\label{eq:intersection-right-flux}\\
 & -\frac{1}{2}d\Omega_{3}\wedge\left(l_{2}dy_{1}^{}\wedge dy_{2}+l_{1}dy_{3}^{}\wedge dy_{4}\right)\nonumber \end{align}
\[
H_{1}=1+\frac{l_{1}}{r^{2}},\quad H_{2}=1+\frac{l_{2}}{r^{2}}.\]
Here $r\equiv\sqrt{z_{1}^{2}+z_{2}^{2}+z_{3}^{2}+z_{4}^{2}}$ is the
radial coordinate in the overall transverse space, $d\Omega_{3}$
is the volume form of the unit radius three-sphere orthogonal to it
in the same space and $l_{1},l_{2}$ are constants proportional to
$g_{s}^{1/2}\alpha'$. The near horizon geometry is given by the limit
$\alpha'\rightarrow0$ with $U=r/\alpha'$ fixed. After this limit
is taken, \eqref{eq:intersection-right-met}\eqref{eq:intersection-right-flux}
becomes\begin{align*}
ds^{2} & =\alpha'\left[\sqrt{L_{1}L_{2}}U^{2}\left(-dw_{0}^{2}+dw_{1}^{2}\right)+\sqrt{L_{1}L_{2}}\frac{dU^{2}}{U^{2}}+\sqrt{L_{1}L_{2}}d\Omega_{3}^{2}\right]\\
 & \quad+\frac{1}{\alpha'\sqrt{L_{1}L_{2}}}\left(dy_{1}^{2}+dy_{2}^{2}+dy_{3}^{2}+dy_{4}^{2}\right),\\
F_{5} & =-\frac{1}{2}Udw_{0}\wedge dw_{1}\wedge dU\wedge\left(dy_{1}\wedge dy_{2}+dy_{3}\wedge dy_{4}\right)-\frac{1}{2}d\Omega_{3}\wedge\left(dy_{1}\wedge dy_{2}+dy_{3}\wedge dy_{4}\right).\end{align*}
where $L_{1,2}\equiv l_{1,2}/\alpha'$ and we have redefined $U\rightarrow\sqrt{L_{1}L_{2}}U,y_{1,2}\rightarrow y_{1,2}/\sqrt{\alpha'L_{2}},\ y_{3,4}\rightarrow y_{3,4}/\sqrt{\alpha'L_{1}}.$
These expressions coincide with \eqref{eq:reduced-met3}\eqref{eq:reduced-flux3}
under the identifications $z=1/U,\ p=\alpha'^{-2}(L_{1}L_{2})^{-1}$.
This near horizon geometry has 16 supersymmetries and thus it can
be seen that, in the case of the solution \eqref{eq:reduced-met2}\eqref{eq:reduced-flux2},
we have 12 enhanced supersymmetries in addition to the 4 supersymmetries
obtained in the previous section. In the case of other solutions for
the Liouville equation, those enhanced symmetries may be inconsistent
with the global identifications in the upper half plane $\mathrm{Im}\xi>0$,
and therefore we expect that the geometries produced by generic solutions
are less supersymmetric than the above geometry produced by a solution
covering the whole of the upper half plane. 

Since the geometries described by \eqref{eq:reduced-met1.5}\eqref{eq:reduced-flux1.5}\eqref{eq:sol-for-v2}\eqref{eq:liouville-eq}
have turned out to be equivalent to the near horizon geometries of
intersecting D3-brane systems, it is valuable to present here the
expressions for some T-dual geometries in our coordinate system. First
we write down the expression for a gauge potential $A_{4}$ of the
five form flux $F_{5}$ in \eqref{eq:reduced-flux1.5}. Noting \eqref{eq:sol-for-v2}\eqref{eq:liouville-eq}
(or \eqref{eq:reduced-dv}), we obtain a solution of the equation
$F_{5}=dA_{4}$ \begin{equation}
A_{4}=\frac{1}{4}dt\wedge V\wedge\left(dy_{1}^{}\wedge dy_{2}^{}+dy_{3}^{}\wedge dy_{4}^{}\right)-\frac{1}{2}\left(dy_{1}^{}\wedge dy_{2}^{}+dy_{3}^{}\wedge dy_{4}^{}\right)\wedge O\label{eq:a4}\end{equation}
where $O$ is a two-form potential of $d\tilde{\Omega}_{3}$. 

To take T-duals of \eqref{eq:reduced-met1.5}\eqref{eq:a4}, we need
the value of the dilaton $\phi_{\mathrm{IIB}}$. We set it equal to
$0$ because in that case we do not have to care about the difference
between Einstein frame and string frame for the above geometries.
Compactifying $R^{4}$ to $T^{4}$ and taking T-dual in the direction
$y_{1}$, we obtain the following type IIA geometries in string frame.
\begin{align}
ds^{2} & =-\frac{1}{\sqrt{p}}\left(dt+V\right)^{2}+\sqrt{p}T^{2}\left(dx_{1}^{2}+dx_{2}^{2}\right)+\frac{1}{\sqrt{p}}dy_{1}^{2}+\sqrt{p}\left(dy_{2}^{2}+dy_{3}^{2}+dy_{4}^{2}\right)+\frac{1}{\sqrt{p}}d\tilde{\Omega}_{3}^{2}\label{eq:near-d2-d4-met}\\
A_{3} & =dt\wedge V\wedge dy_{2}^{}-2dy_{2}\wedge O,\qquad\phi_{\mathrm{IIA}}=-\frac{1}{4}\ln p\label{eq:near-d2-d4-flux}\\
 & V_{}=\frac{1}{4}\epsilon_{ij}\partial_{j}\ln T^{2}dx^{i},\qquad\left(\partial_{1}^{2}+\partial_{2}^{2}\right)\ln\left(T(x)^{2}\right)=8pT(x)^{2}.\nonumber \end{align}
These are locally equivalent to the near horizon geometries of D2-D4
systems. Further, taking T-dual in the direction $y_{2}$, we obtain
the following type IIB geometries in string frame.\begin{align}
ds^{2} & =-\frac{1}{\sqrt{p}}\left(dt+V\right)^{2}+\sqrt{p}T^{2}\left(dx_{1}^{2}+dx_{2}^{2}\right)+\frac{1}{\sqrt{p}}\left(dy_{1}^{2}+dy_{2}^{2}\right)+\sqrt{p}\left(dy_{3}^{2}+dy_{4}^{2}\right)+\frac{1}{\sqrt{p}}d\tilde{\Omega}_{3}^{2}\label{eq:near-d1-d5-met}\\
A_{2} & =dt\wedge V-2O,\qquad\phi_{\mathrm{IIB}}=-\frac{1}{2}\ln p\label{eq:near-d1-d5-flux}\\
 & V_{}=\frac{1}{4}\epsilon_{ij}\partial_{j}\ln T^{2}dx^{i},\qquad\left(\partial_{1}^{2}+\partial_{2}^{2}\right)\ln\left(T(x)^{2}\right)=8pT(x)^{2}.\nonumber \end{align}
These are locally equivalent to the near horizon geometries of frequently-discussed
D1-D5 systems.

\subsubsection*{\emph{Wick rotation}}

Finally, since we have understood the basic properties of the geometries
with the $S^{3}\times S^{3}$ factor, we comment on the possibility
that there is a connection to the results of other works. Liouville
theory has appeared also in other contexts as in \cite{Coussaert:1995zp,Seiberg:1999xz}.
Our result is different from theirs in that \eqref{eq:liouville-eq}
is a Euclidean Liouville equation. Interestingly, as we will see below,
we can find analytic continuations to Minkowskian Liouville equations
such that the resultant metrics and fluxes are again $AdS_{3}\times S^{3}\times R^{4}$
solutions of the same supergravity. 

Let us recall the form of the general solution of the Liouville equation
\eqref{eq:general-sol}. We can always take $\xi=x'_{1}+ix'_{2}$
as coordinates of $(x_{1},x_{2})$ space, and because $\xi(u)$ is
a holomorphic function, the conformal form of the metric in two-dimensional
space $g_{ij}\sim T^{2}\delta_{ij}$ is not affected by this coordinate
transformation. Thus we see that the metric and five-form flux expressed
in the coordinate $\xi$ are also in our ansatz and satisfy the same
constraints. The difference from \eqref{eq:reduced-met2}\eqref{eq:reduced-flux2}
is just that $x_{1}$ and $x_{2}$ are interchanged and some signs
are flipped. Explicitly, the metric and five-form flux are given by
\begin{align*}
ds^{2} & =-\frac{1}{\sqrt{p}}\left(dt+V\right)^{2}+\frac{1}{4\sqrt{p}x{}_{2}^{\prime2}}\left(dx_{1}^{\prime2}+dx_{2}^{\prime2}\right)+\sqrt{p}\left(dy^{2}+y^{2}d\hat{\Omega_{}}_{3}^{2}\right)+\frac{1}{\sqrt{p}}d\tilde{\Omega}_{3}^{2}\\
F_{5} & =-\frac{1}{4}dt\wedge dV\wedge\left(dy_{1}^{}\wedge dy_{2}^{}+dy_{3}^{}\wedge dy_{4}^{}\right)-\frac{1}{2}\left(dy_{1}^{}\wedge dy_{2}^{}+dy_{3}^{}\wedge dy_{4}^{}\right)\wedge d\tilde{\Omega}_{3}\\
V & =-\frac{1}{2x'_{2}}dx'_{1}.\end{align*}
From this we see that the Wick-rotations $x_{1}'\rightarrow\pm ix_{1}'$
lead to that $V$ turns into $\pm iV$, $T^{2}=1/4px_{2}^{\prime2}$
is unchanged and \eqref{eq:liouville-eq} turns into a Minkowskian
Liouville equation. To keep the metric real, we need another Wick
rotation, $t\rightarrow\pm it$. After this double Wick rotation,
the five-form flux remains real, and hence these metric and the flux
are again a solution of IIB supergravity. The coordinate transformation
in the three dimensional subspace which leads to the Poincar$\Acute{\mathrm{e}}$
metric (In the case of \eqref{eq:reduced-met2}\eqref{eq:reduced-flux2}
it was \eqref{eq:coord-trans-to-poincare}.) can be used with the
corresponding Wick rotations $x^{\prime+,-}\rightarrow\pm ix^{\prime+,-}$.
Thus we see that this Wick rotated geometry is again an $AdS_{3}\times S^{3}\times R^{4}$
solution and it is the near horizon geometry of an intersecting D3
brane system.

The above Wick rotations work also for the cases of T-dualized geometries
\eqref{eq:near-d2-d4-met} \eqref{eq:near-d2-d4-flux} \eqref{eq:near-d1-d5-met}
\eqref{eq:near-d1-d5-flux}. The appearance of Liouville theory for
every slice of constant $t$ may lead to some understanding of D1-D5
systems in the future.

\section{Conclusion\label{sec:Conclusion}}

In this paper, we have shown that a new differential equation should
be imposed on the resultant controlling functions $m,n,p,T$ of \cite{Gava:2006pu},
and discussed the limit $n=0,m=1/y^{2}$, in which the new equation
plays crucial roles to obtain some properties of the geometries. 

Among the properties recognized in this paper, the appearance of Liouville
theory in D1-D5 systems seems to be related most directly to other
works. In \cite{Coussaert:1995zp,Seiberg:1999xz}, it has been shown
that some boundary dynamics in $AdS_{3}$ are related to Liouville
theory. In \cite{Boni:2005sf}, properties possessed by the solutions
of Liouville equation have been found in supergravity. Investigating
the precise relations of our result to those works will be our next
task. The interesting point is that, in contrast to them, we have
obtained Liouville equation itself in the bulk.

Next of interest is in the possibility that our results may relate
the spectra of different conformal field theories. The world volume
theories on intersecting D3-branes have been constructed in \cite{Constable:2002xt}
and those of D1-D5 systems are often discussed. Although we have not
investigated how the global structures of the two-dimensional surfaces
described by the Liouville theory are related to those of the ten-dimensional
geometries, we have shown that those structures are common to the
T-dualized geometries. This implies that, if there exists a solution
which is regarded as an excited state in the near horizon geometry
of one configuration of D-branes, we can map it to those of T-dualized
systems. This kind of duality relation in the gravity sides may lead
to new understandings about the relations between the two different
dual field theories. 

Another of interest is in the relation between $AdS_{3}\times S^{3}\times R^{4}$
and $AdS_{5}\times S^{5}$. In LLM, $AdS_{5}\times S^{5}$ is described
by a circular droplet with its radius equal to the square of that
of the $AdS_{5}$, and the limit of large radius or small radius corresponds
to the limit $n=p=0,m=1/y^{2}$ in our geometries, which can also
be considered as a large radius limit of $AdS_{3}$. More generally,
for any droplet configuration of LLM, this geometry can be obtained
by looking closely around any point on the $y=0$ plane. Although
this is a singular geometry and the validity of supergravity approximation
has to be discussed, it might imply something new about the behaviors
of $\mathcal{N}=4$ SYM in the corresponding limits. $ $

\subsection*{Acknowledgements}

I specially thank T. Yoneya for much advice during this work and many
instructive comments on the manuscript. I would also like to thank
Y. Aisaka, T. Matsuda, A. Miwa, A. Tsuji and N. Yokoi for useful discussions.
This work was supported in part by JSPS Research Fellowships for Young
Scientists.

\appendix

\section{Building blocks of squashed three-spheres\label{sec:Squashed-three-sphere}}

In this section, we explain the relation between the unit vector in
four-dimensional space used in Sec.\ref{sec:s3xs3} and building blocks
used for defining the metrics of squashed three-spheres.

Let us parametrize the unit vector in four-dimensional space as follows.\begin{eqnarray*}
\hat{y}_{1} & = & \cos\frac{\theta}{2}\cos\frac{\psi+\phi}{2}\\
\hat{y}_{2} & = & -\cos\frac{\theta}{2}\sin\frac{\psi+\phi}{2}\\
\hat{y}_{3} & = & -\sin\frac{\theta}{2}\cos\frac{\psi-\phi}{2}\\
\hat{y}_{4} & = & \sin\frac{\theta}{2}\sin\frac{\psi-\phi}{2}.\end{eqnarray*}
 In this parametrization, the metric of $S^{3}$ is\begin{eqnarray}
ds^{2} & = & d\hat{y}_{1}^{2}+d\hat{y}_{2}^{2}+d\hat{y}_{3}^{2}+d\hat{y}_{4}^{2}\nonumber \\
 & = & \frac{1}{4}\left(d\theta^{2}+d\phi^{2}+d\psi^{2}+2\cos\theta d\phi d\psi\right).\label{eq:s3-met}\end{eqnarray}

To define the metric of squashed three-spheres, we consider $SU(2)_{L}$
and $SU(2)_{R}$ generators,

\begin{align}
L^{1}=\left(\begin{array}{cccc}
 & -1\\
1\\
 &  &  & 1\\
 &  & -1\end{array}\right), & L^{2}=\left(\begin{array}{cccc}
 &  & -1\\
 &  &  & -1\\
1\\
 & 1\end{array}\right),L^{3}=\left(\begin{array}{cccc}
 &  &  & 1\\
 &  & -1\\
 & 1\\
-1\end{array}\right),\label{eq:def-of-ls}\\
R^{1}=\left(\begin{array}{cccc}
 & -1\\
1\\
 &  &  & -1\\
 &  & 1\end{array}\right), & R^{2}=\left(\begin{array}{cccc}
 &  & -1\\
 &  &  & 1\\
1\\
 & -1\end{array}\right),R^{3}=\left(\begin{array}{cccc}
 &  &  & -1\\
 &  & -1\\
 & 1\\
1\end{array}\right).\label{eq:def-of-rs}\end{align}
We can construct left-invariant one-forms from $SU(2)_{R}$ generators\begin{align}
R_{\alpha\beta}^{1}\hat{y}_{\alpha}^{}d\hat{y}_{\beta} & =-\sigma_{\hat{3}}\label{eq:r1-sigma3}\\
R_{\alpha\beta}^{2}\hat{y}_{\alpha}^{}d\hat{y}_{\beta} & =-\sigma_{\hat{1}}\nonumber \\
R_{\alpha\beta}^{3}\hat{y}_{\alpha}d\hat{y}_{\beta} & =\sigma_{\hat{2}}.\nonumber \end{align}
The metrics of $SU(2)_{L}$ invariant squashed three-spheres are given
by\[
ds^{2}=r_{ij}\left(R_{\alpha\beta}^{i}\hat{y}_{\alpha}^{}d\hat{y}_{\beta}\right)\left(R_{\gamma\delta}^{j}\hat{y}_{\gamma}^{}d\hat{y}_{\delta}\right)\]
where $r_{ij}^{}$ is an arbitrary symmetric tensor. We can define
the metrics of $SU(2)_{R}$ invariant squashed three-spheres by using
the $SU(2)_{L}$ generators \eqref{eq:def-of-ls}.\[
ds^{2}=l_{ij}\left(L_{\alpha\beta}^{i}\hat{y}_{\alpha}^{}d\hat{y}_{\beta}\right)\left(L_{\gamma\delta}^{j}\hat{y}_{\gamma}d\hat{y}_{\delta}\right).\]
If $r_{ij}=l_{ij}=\delta_{ij}$, the two metrics coincide and are
equal to \eqref{eq:s3-met}, and each symmetry is enhanced to $SO(4)=SU(2)_{L}\times SU(2)_{R}$. 

\bibliographystyle{utphys}
\addcontentsline{toc}{section}{\refname}\bibliography{brane,bps-sol,2d-gravity,d1-d5,duality,giant-graviton}

\providecommand{\href}[2]{#2}\begingroup\raggedright\begin{thebibliography}{10}

\bibitem{Lin:2004nb}
H.~Lin, O.~Lunin, and J.~M. Maldacena, ``Bubbling ads space and 1/2 bps
  geometries,'' {\em JHEP} {\bf 10} (2004) 025,
\href{http://www.arXiv.org/abs/hep-th/0409174}{{\tt hep-th/0409174}}.

\bibitem{Kim:2005ez}
N.~Kim, ``{AdS(3) solutions of IIB supergravity from D3-branes},'' {\em JHEP}
  {\bf 01} (2006) 094,
\href{http://www.arXiv.org/abs/hep-th/0511029}{{\tt hep-th/0511029}}.

\bibitem{Donos:2006iy}
A.~Donos, ``{A description of 1/4 BPS configurations in minimal type IIB
  SUGRA},'' {\em Phys. Rev.} {\bf D75} (2007) 025010,
\href{http://www.arXiv.org/abs/hep-th/0606199}{{\tt hep-th/0606199}}.

\bibitem{Donos:2006ms}
A.~Donos, ``{BPS states in type IIB SUGRA with SO(4) x SO(2)(gauged)
  symmetry},'' {\em JHEP} {\bf 05} (2007) 072,
\href{http://www.arXiv.org/abs/hep-th/0610259}{{\tt hep-th/0610259}}.

\bibitem{Gava:2006pu}
E.~Gava, G.~Milanesi, K.~S. Narain, and M.~O'Loughlin, ``1/8 bps states in
  ads/cft,'' {\em JHEP} {\bf 05} (2007) 030,
\href{http://www.arXiv.org/abs/hep-th/0611065}{{\tt hep-th/0611065}}.

\bibitem{Chen:2007du}
B.~Chen {\em et al.}, ``{Bubbling AdS and droplet descriptions of BPS
  geometries in IIB supergravity},'' {\em JHEP} {\bf 10} (2007) 003,
\href{http://www.arXiv.org/abs/arXiv:0704.2233 [hep-th]}{{\tt arXiv:0704.2233
  [hep-th]}}.

\bibitem{Lunin:2008tf}
O.~Lunin, ``{Brane webs and 1/4-BPS geometries},''
\href{http://www.arXiv.org/abs/arXiv:0802.0735 [hep-th]}{{\tt arXiv:0802.0735
  [hep-th]}}.

\bibitem{Youm:1997hw}
D.~Youm, ``Black holes and solitons in string theory,'' {\em Phys. Rept.} {\bf
  316} (1999) 1--232,
\href{http://www.arXiv.org/abs/hep-th/9710046}{{\tt hep-th/9710046}}.

\bibitem{Coussaert:1995zp}
O.~Coussaert, M.~Henneaux, and P.~van Driel, ``The asymptotic dynamics of
  three-dimensional einstein gravity with a negative cosmological constant,''
  {\em Class. Quant. Grav.} {\bf 12} (1995) 2961--2966,
\href{http://www.arXiv.org/abs/gr-qc/9506019}{{\tt gr-qc/9506019}}.

\bibitem{Seiberg:1999xz}
N.~Seiberg and E.~Witten, ``The d1/d5 system and singular cft,'' {\em JHEP}
  {\bf 04} (1999) 017,
\href{http://www.arXiv.org/abs/hep-th/9903224}{{\tt hep-th/9903224}}.

\bibitem{Boni:2005sf}
M.~Boni and P.~J. Silva, ``Revisiting the d1/d5 system or bubbling in ads(3),''
  {\em JHEP} {\bf 10} (2005) 070,
\href{http://www.arXiv.org/abs/hep-th/0506085}{{\tt hep-th/0506085}}.

\bibitem{Constable:2002xt}
N.~R. Constable, J.~Erdmenger, Z.~Guralnik, and I.~Kirsch, ``Intersecting
  d3-branes and holography,'' {\em Phys. Rev.} {\bf D68} (2003) 106007,
\href{http://www.arXiv.org/abs/hep-th/0211222}{{\tt hep-th/0211222}}.

\end{thebibliography}\endgroup

\end{document}